\newif\ifARXIV
\ARXIVtrue

\ifARXIV
\documentclass[sigconf,nonacm]{acmart}
\else
\documentclass[sigconf]{acmart}
\fi

\usepackage{listings}
\usepackage{float}

\newif\ifBIBLATEX
\BIBLATEXfalse

\newif\ifDEBUG
\DEBUGtrue
\DEBUGfalse

\ifBIBLATEX
  \usepackage[style=numeric,
  maxnames=200,
  backend=biber,
  firstinits=true]{biblatex}
\fi

\usepackage{multirow}

\usepackage{comment}

\usepackage{amsmath,amsfonts}
\usepackage{algorithmic}
\usepackage{graphicx}
\usepackage{textcomp}
\usepackage{xcolor}
\usepackage{booktabs} 
\usepackage{xspace} 
\usepackage[normalem]{ulem}
\usepackage{makecell}
\usepackage{tcolorbox}
\usepackage{enumitem}
\usepackage{siunitx}
\usepackage[vskip=1em,font=itshape,leftmargin=2em,rightmargin=2em]{quoting}

\usepackage{soul}

\ifDEBUG
    \newcommand{\JD}[1]{\textcolor{purple}{[JD:#1]}}
    \newcommand{\TRS}[1]{\textcolor{olive}{[Taylor:#1]}}
    
\else
    \newcommand{\JD}[1]{}
    \newcommand{\TRS}[1]{}

\fi

\newif\ifSPACEHACK
\SPACEHACKfalse
    \usepackage{etoolbox}
    \makeatletter
    \patchcmd{\ttlh@hang}{\parindent\z@}{\parindent\z@\leavevmode}{}{}
    \patchcmd{\ttlh@hang}{\noindent}{}{}{}
    \makeatother

\ifSPACEHACK
    \titlespacing*\section{0pt}{1pt plus 1pt minus 1pt}{1pt plus 1.5pt minus 1.5pt}
    \titlespacing*\subsection{0pt}{1pt plus 1.5pt minus 1.5pt}{1pt plus 1.5pt minus 1.5pt}
    \titlespacing*\subsubsection{0pt}{2pt plus 1pt minus 1pt}{1pt plus 1.5pt minus 1.5pt}
    \titlespacing*\paragraph{0pt}{1pt plus 1.5pt minus 1.5pt}{1pt plus 1.5pt minus 1.5pt}
\fi
\usepackage{balance} 
\usepackage{subcaption}

\usepackage{amsmath}
\usepackage{cleveref}

\crefformat{section}{\S#2#1#3}
\crefname{figure}{Figure}{Figures}
\crefname{appendix}{Appendix}{Appendices}
\crefname{table}{Table}{Tables}
\crefname{algorithm}{Algorithm}{Algorithms}
\crefname{listing}{Listing}{Listings}
\crefname{theorem}{Theorem}{Theorems}
\crefname{thm}{Theorem}{Theorems}
\crefname{lemma}{Lemma}{Lemmata}
\crefname{equation}{Eqt.}{Eqts.}
\crefformat{Grammar}{Grammar #1}

\usepackage{enumitem}
\usepackage{booktabs}

\newcommand{\ie}{\textit{i.e.,} }
\newcommand{\eg}{\textit{e.g.,} }
\newcommand{\etal}{\textit{et al.}\xspace}

\newcommand{\code}[1]{\emph{#1}\xspace}

\copyrightyear{2022}
\acmYear{2022}
\setcopyright{rightsretained}
\acmConference[SCORED '22]{Proceedings of the 2022 ACM Workshop on Software Supply Chain Offensive Research and Ecosystem Defenses}{November 11, 2022}{Los Angeles, CA, USA}
\acmBooktitle{Proceedings of the 2022 ACM Workshop on Software Supply Chain Offensive Research and Ecosystem Defenses (SCORED '22), November 11, 2022, Los Angeles, CA, USA}
\acmDOI{10.1145/3560835.3564556}
\acmISBN{978-1-4503-9885-5/22/11}

\acmSubmissionID{scor023}

\usepackage{etoolbox}
\makeatletter
\patchcmd{\maketitle}{\@copyrightpermission}{
   \begin{minipage}{0.3\columnwidth}
     \href{https://creativecommons.org/licenses/by/4.0/}{\includegraphics[width=0.90\textwidth]{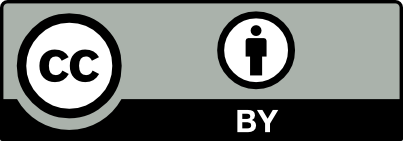}}
   \end{minipage}\hfill
   \begin{minipage}{0.7\columnwidth}
     \href{https://creativecommons.org/licenses/by/4.0/}{This work is licensed under a Creative Commons Attribution International 4.0 License.}
   \end{minipage}

   \vspace{5pt}
}{}{}

\makeatother

\begin{document}

\begin{abstract}
This paper systematizes knowledge about secure software supply
chain patterns. It identifies four stages of a software supply chain
attack and proposes three security properties crucial for a secured supply
chain: transparency, validity, and separation. The paper describes
current security approaches and maps them to the proposed security
properties, including research ideas and case studies of supply
chains in practice. It discusses the strengths and weaknesses of current
approaches relative to known attacks and details the various security frameworks put out to ensure the security of the software supply chain. Finally, the paper highlights potential gaps in actor and operation-centered supply chain security techniques.
\end{abstract}




\author{Chinenye Okafor}
\authornote{Both authors contributed equally to this research.}
\orcid{0000-0002-4853-6870}
\affiliation{%
  \institution{Purdue University}
  \country{West Lafayette, IN, USA}}
\email{okafor1@purdue.edu}

\author{Taylor R. Schorlemmer}
\authornotemark[1]
\orcid{0000-0003-2181-5527}
\affiliation{%
  \institution{Purdue University}
  \country{West Lafayette, IN, USA}}
\email{tschorle@purdue.edu}

\author{Santiago Torres-Arias}
\orcid{0000-0002-9283-3557}
\affiliation{%
  \institution{Purdue University}
  \country{West Lafayette, IN, USA}}
\email{santiagotorres@purdue.edu}

\author{James C. Davis}
\orcid{0000-0003-2495-686X}
\affiliation{%
  \institution{Purdue University}
  \country{West Lafayette, IN, USA}}
\email{davisjam@purdue.edu}

\renewcommand{\shortauthors}{Chinenye Okafor, Taylor R. Schorlemmer, Santiago Torres-Arias, \& James C. Davis}

\ccsdesc[300]{Software and its engineering~Design patterns}
\ccsdesc[300]{Security and privacy~Software security engineering}
\ccsdesc[300]{General and reference~Surveys and overviews}

\keywords{Software Supply Chain Attacks, Security Properties, Collaborative Software Engineering, Software Reuse}

\title{SoK: Analysis of Software Supply Chain Security by Establishing Secure Design Properties}

\maketitle

\section{Introduction}


Industrial, government, and academic computing systems rely on a supply chain of open- and closed-source software components~\cite{raymond1999cathedral}.
An actor controlling any step in this chain may, accidentally or maliciously, sabotage downstream software~\cite{prana2021out,sejfia2022practical}.
Problems in software supply chains have caused site- and Internet-wide disruptions at an estimated cost of billions of dollars~\cite{ohm2020backstabber,notpetya-attack}.
These problems include service outages~\cite{leftpad,dellavecchiaHowRogueDeveloper2022} and cybersecurity exploits that endanger human lives~\cite{forbes2022supplychain} and national security~\cite{solarwinds}.
Software supply chain exploits may be attributed to requirements mismatch --- many software supply chains were originally designed for sharing, not cybersecurity~\cite{raymond1999cathedral}.
In light of the emerging requirement for security, how should supply chains be designed?

Researchers have begun by structuring knowledge of how current software supply chains can be attacked.
For example, taxonomies of attacks yield attack trees that help us understand how attackers compromise supply chains.
Ecosystem analysis has also helped understand how the structure of software dependencies can make us more or less vulnerable when selecting external dependencies~\cite{Zimmermann2019SecurityThreatsinNPMEcosystem,Decan2018SecurityVulnerabilitiesinNPMDependencyNetwork,zerouali2022impact}.
Data-science based efforts in the industry and academia have attempted to identify signals or indicators of compromise~\cite{Zahan2022WeakLinksinNPMSupplyChain}. 

Other researchers have examined design changes to improve the security of software supply chains.
Using these insights, efforts have focused in developing systems and mechanisms to mitigate these attack vectors. 
Efforts like in-toto and Sigstore attempt to provide a layer of security to the current operations in the software supply chain. 
These generally aim to protect against a class of attacks in the software supply chain.
For example, Solarwind's Trebuchet project~\cite{solarwinds} aims to prevent compiled backdoors by means of an in-toto coordinated reproducible builds-based pipeline.
This has lead to the development of ``meta-frameworks'' or ``best practices models'' that describe a combination of mechanisms and configurations that can be used to provide a strong security posture against software supply chain attacks.
Examples of these approaches are the Cloud Native Computing Foundations Technical Advisory Group on Security's (CNCF TAG-Security) reference architecture for secure software pipelines~\cite{cloud_native_computing_foundation_software_2021}, as well as the Secure Software Factory~\cite{security_technical_advisory_group_secure_2022}.

Given the emerging nature of this discipline and the amount of disjoint efforts, various groups from academia, industry and open source have proposed multiple reference architectures and design patterns to secure their software supply chain.
However, as yet there is no a systematized framework that helps system integrators and designers to understand and match how these mechanisms and different architectures are designed and applied in popular software pipelines.
This is in part due to the disconnect between these different communities, but as well as a lack of structured framework to map both research aims: to catalogue threats and to apply a combination of systems to mitigate them.


The purpose of this paper is to summarize knowledge regarding design patterns for security in software supply chains.
We systematically review current design patterns for secure supply chains, and develop a framework to compare their security postures. In doing so, we provide the first comprehensive study of current best practices proposed by industry, academia and government.
First, we describe software supply chains (\cref{sec:background}).
Next, we provide a four stage attack pattern for software supply chain attacks (\cref{sec:Attacks}).
Then, we present three properties (transparency, validity, and separation) for securing software supply chains (\cref{sec:properties}).
Afterwards, we document how current security practices fulfill our security principles (\cref{sec:approaches}) and provide several case studies (\cref{sec:embodiments}).
Finally, we identify opportunities to further improve the security of software supply chains (\cref{sec:discussion}).


\begin{figure*}[h!!]
    \centering
    \includegraphics[width=0.8\textwidth]{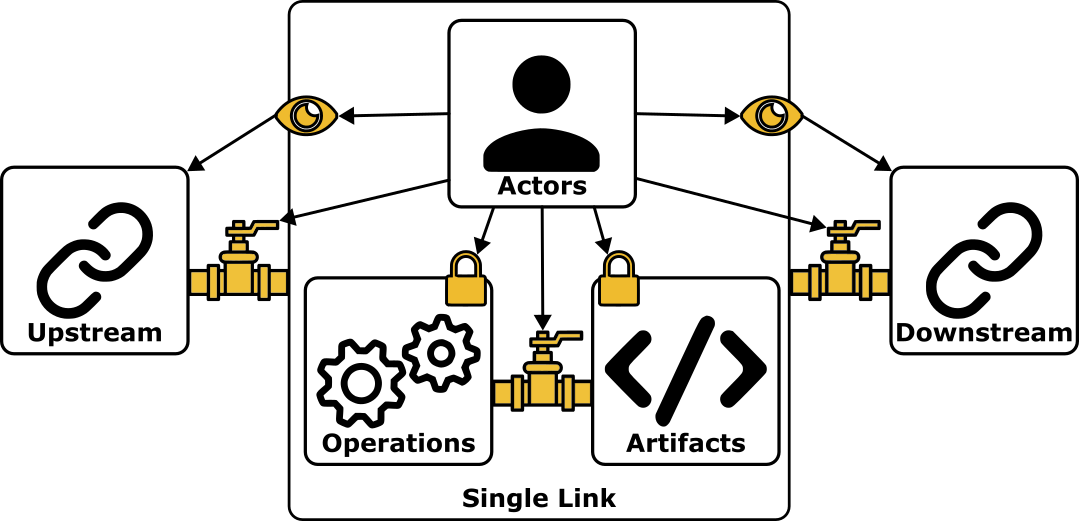}
    \caption{A software supply chain with focus on a single link. Actors manage components and connections within and between links. Therefore, actors manage security. Security depends on upstream and downstream transparency, link validity via component integrity and actor authentication, and logical separation between components and links.}
    \label{fig:supply_chain}
\end{figure*}

In summary, our contributions are:
\begin{enumerate}
    \item An attack pattern for software supply chain attacks (\cref{sec:Attacks}).
    \item Principles for a secure software supply chain (\cref{sec:properties}). 
    \item A collection and analysis of current security practices (\cref{sec:approaches} and \cref{sec:embodiments}).
\end{enumerate}

\section{Background} \label{sec:background}

In general, a supply chain is a set of entities which interact to produce some product for an end consumer \cite{stadtler_supply_2008}.
Each link in a supply chain contributes to the final product by providing a sub-product to reliant links.
As a result, a network of dependencies forms between the links in a supply chain - terminating with the link representing the end consumer.
Therefore, a supply chain is characterized by the connections and attributes of the entities used to create and ultimately consume a final product.

In computing, \textbf{software supply chains are a collection of systems, devices, and people which produce a final software product}~\cite{cybersecurity_enisa_2021}. 
\cref{fig:supply_chain} depicts a typical software supply chain with a focus on an individual link.
Each \emph{link} in the software supply chain comprises the artifacts, operations, and actors needed to develop and deliver software products \cite{ellison2010evaluating,nissen2018deliver}.
\emph{Actors} manipulate artifacts and operations within the supply chain to produce an output.
\emph{Artifacts} include the product team's code, development infrastructure, and software dependencies.
\emph{Operations} include
  productive steps such as fetching dependencies or compiling software,
  protective steps such as linting or security scans,
  and
  publishing steps such as deployment or distribution.

The structure of supply chains necessitates an interdependence between artifacts and operations within and between links. 
Actors manage the connections which form between components and between links in the chain.
Responsibility for operations and artifacts is distributed among actors across different geographies, teams, companies, and legal jurisdictions.
Modern software engineering is an international collaborative effort \cite{herbsleb2007global,vsmite2010empirical}.
A single link in the supply chain does not necessarily correspond to one group or organization. A single organization may be responsible for several links within a supply chain.

\section{Supply Chain Attacks and Security Properties} \label{sec:AttacksAndProperties}

\subsection{Supply Chain Attacks}  \label{sec:Attacks}

\begin{figure}
    \centering
    \includegraphics[width=0.4\textwidth]{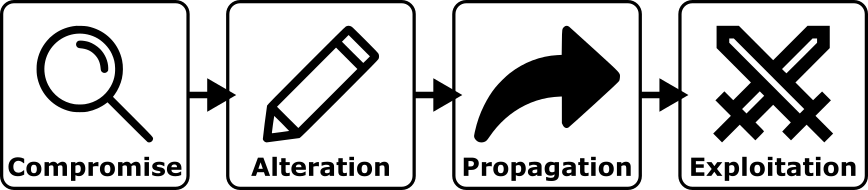}
    \caption{Four stage software supply chain attack pattern. Attackers begin with an initial \textit{compromise} before making some malicious \textit{alteration} to the supply chain. This change then \textit{propagates} down the supply chain where attackers \textit{exploit} the introduced weakness(es).}
    \label{fig:attack}
\end{figure}

The software supply chain is an increasingly popular attack vector \cite{Ladisa2022TaxonomyofAttacksonOSSSupplyChains}.
It is comprised of several connected links which share artifacts and conduct operations. Actors manage links and components.
The difference between software supply chain attacks and other software attacks, however, is not clearly defined in literature.
Ladisa \etal\cite{Ladisa2022TaxonomyofAttacksonOSSSupplyChains} and Ohm \etal \cite{ohm2020backstabber} characterize supply chain attacks as the injection of malicious code into the supply chain to target downstream links.
ENISA \cite{cybersecurity_enisa_2021} defines a supply chain attack as a combination of at least two attacks --- one attack on a supplier and a subsequent attack on intended targets.
Other works such as Zimmermann \etal\cite{Zimmermann2019SecurityThreatsinNPMEcosystem} and Zahan \etal\cite{Zahan2022WeakLinksinNPMSupplyChain} identify methods other than strict code injection for supply chain attacks.
\textbf{Distilling the concept of software supply chain attacks from multiple sources, we arrive at a characteristic four stage attack pattern shown in~\cref{fig:attack}}:

\begin{enumerate}
\item \textbf{Compromise:}
First, an attacker finds and compromises an existing weakness within a supply chain. 

\item \textbf{Alteration:}
Second, an attacker leverages the initial compromise to alter the software supply chain.

\item \textbf{Propagation:}
Third, the change introduced by the attacker propagates to downstream components and links. 

\item \textbf{Exploitation:}
Finally, the attacker exploits the alterations in a downstream link.
\end{enumerate}

To illustrate this definition, consider the SOLARBURST compromise~\cite{solarwinds-microsoft,solarwinds-zdnet}.
In this supply chain attack, an attacker altered existing software from SolarWinds by injecting malicious code during the build process.
This attack can be mapped to the four-stage attack from~\cref{fig:attack} as follows:
  (1) The \emph{existing weakness} compromised was the build infrastructure.
  (2) The \emph{alteration} was malicious code injected by the compiler, permitting a user to bypass authentication in a SolarWinds product component.
  (3) The \emph{propagation} was via SolarWinds's compromised product --- its users include many companies and US government agencies like the IRS and NASA.
  (4) The \emph{exploitation} was to leverage broken authentication mechanism to take control of affected machines.
Such incidents are becoming common; software supply chain compromises have increased by a cumulative 650\% in the last three years~\cite{sonatype-sotssc, istr-2018, istr-2019}.

In contrast to this pattern, traditional attacks, such as those described by Lockheed Martin's Cyber Kill Chain \cite{lockheed_martin_cyber_2022}, simply exploit an existing vulnerability (step 4).
Attacks on software are not necessarily supply chain attacks just because the software exists within the context of a supply chain.
For this reason, \emph{an attack on software via the weakness of a dependency is not a supply chain attack unless it follows the attack pattern}; the attacker must both introduce the upstream change \emph{and} subsequently exploit it downstream.

For another perspective on this attack pattern, consider the distinction between vulnerable and malicious dependencies as drawn by the European Union Agency for Cybersecurity (ENISA)~\cite{cybersecurity_enisa_2021} and Ohm \etal~\cite{ohm2020backstabber}.
Vulnerable links in the supply chain contain unintended weaknesses that may be exploited further downstream.
These exploits are not supply chain attacks.
On the other hand, malicious links in the supply chain were intentionally designed to weaken the rest of the chain.
Introducing and subsequently exploiting such weaknesses constitutes a supply chain attack.

Existing literature categorizes and documents known supply chain attacks \cite{ohm2020backstabber, Ladisa2022TaxonomyofAttacksonOSSSupplyChains, Zimmermann2019SecurityThreatsinNPMEcosystem, cybersecurity_enisa_2021}.
It is outside of the scope of this paper to enumerate individual attack types.
Typically, this line of work differentiates between how attackers compromise and alter the supply chain.




\subsection{Security Properties for Software Supply Chains} \label{sec:properties}

Components of a supply chain must be secured to mitigate the presence of vulnerabilities and the risk of attack. Supply chains become secure when attackers are unable to compromise components, alter the supply chain, or propagate malicious changes.
\textbf{In the literature on software supply chain security, we have identified three orthogonal and recurring security properties}:

\begin{enumerate}
\item \textbf{Transparency:}
Although actors only control portions of a supply chain, increased knowledge of the entire chain allows all parties to mitigate risk or employ specific countermeasures against an attack \cite{microsoft_3_2021, ellison_evaluating_2010, Ladisa2022TaxonomyofAttacksonOSSSupplyChains, ensor_shifting_2021, boyens_cybersecurity_2022}. 
Transparency represents the availability of that knowledge to actors in the supply chain. 
Transparency applies to the entities connecting and comprising links in the chain. 

\item \textbf{Validity:}
Software supply chains should remain correct. Changes to actors, operations, or artifacts in a single link can compromise downstream entities.
Validity comprises integrity of operations, integrity of artifacts, and authentication of actors.
Each link in the supply chain contains a series of operations and artifacts which interact with other links in the chain. Secure supply chains require that these components remain unchanged by malicious parties \cite{microsoft_3_2021, ensor_shifting_2021, boyens_cybersecurity_2022}.
Therefore, only authorized actors should make changes to link connections and components. Such changes must also receive permission to occur \cite{in-toto-usenix, ensor_shifting_2021, Zahan2022WeakLinksinNPMSupplyChain}.



\item \textbf{Separation:}
Secure supply chains embody a compartmentalized nature. 
Connections are an integral part of supply chains, but should only exist when necessary. These connections should be minimized to reduce attack surface area. Additionally, logically separate operations, artifacts, and actors should remain separate in practice to minimize unintended connections.
By implementing measures such as mirroring, version locking, containers etc. individual components can decrease reliance on security of others \cite{microsoft_3_2021, Ladisa2022TaxonomyofAttacksonOSSSupplyChains, Zahan2022WeakLinksinNPMSupplyChain}. 
\end{enumerate}

\subsection{Analysis of Security Properties}

The security properties discussed in \cref{sec:properties} are only meaningful if, when applied ideally, they eliminate the risk of attack. 
For this to be the case, properties must be comprehensive.

To analyze these properties, we consider a hypothetical attack following the pattern discussed in \cref{sec:Attacks} and apply transparency, validity, and separation. 
Since defenders can typically only address the first there stages of attack, we show how applying these properties prevents an attack from reaching the final stage: exploitation.
Lastly, we note the difference between ideal conceptualizations and real-world embodiments of these security properties.

First, transparency primarily protects against the first stage of attack.
Transparency, in its ideal state, enables perfect vision of all actors, operations, and artifacts across the supply chain.
Such transparency would allow managers of a supply chain to identify link weaknesses before they are compromised.
By securing weaknesses through patches, fixes, or other methods, managers block attempts at \textit{compromise}. By identifying weaknesses first, managers prevent attackers from completing the first stage.

Second, validity primarily protects against the next stage of attack: \textit{alteration}.
By maintaining perfect integrity of operations, integrity of artifacts, and authentication of actors, no unauthorized changes can be made to the supply chain.
As a result, attackers have no ability to maliciously alter the supply chain.

Finally, separation primarily protects against the third stage of attack: \textit{propagation}.
If a supply chain system can perfectly compartmentalize and moderate interactions between entities, then malicious changes cannot propagate downstream.
In this case, connections between artifacts, operations, and actors are managed in such a way that malicious changes cannot affect other supply chain components.
With ideal separation, only valid changes (\eg system updates and patches) can traverse the supply chain.

By preventing at least one of the three stages leading to exploitation, a supply chain attack cannot occur.
While this hypothetical considers an ideal case, practical application is not so easy.
Real techniques typically do not fully realize security properties, but they provide partial coverage (\eg attestations do not necessarily provide complete transparency).
Real techniques also do not always implement security properties independently.
In practice, techniques might require cohesion between multiple security properties.
For example, a technique might require transparency to identify threats propagating through the supply chain before applying separation methods.
Conversely, achieving close-to-ideal transparency may only be possible if the supply chain is sufficiently separated from other entities.

Since techniques do not perfectly embody security properties, defending in depth is critical~\cite{joint_task_force_interagency_working_group_security_2020}.
Theoretically, a single technique could prevent all attacks if it provided a perfect implementation of transparency, validity, or separation.
In practice, techniques have flaws.
For this reason, using multiple techniques mapped to each of the security properties provides a more effective defense against attack.

\section{Mapping Proposals to Security Properties} \label{sec:approaches}

In this section, we map proposals to secure the software supply chain against the security properties presented in~\cref{sec:properties}.

\subsection{Promoting Transparency}

A lot of work has been done on creating transparency in the software supply chain. To mitigate the security risks associated, it is crucial to have information about the various software components and dependency as well as their hierarchical relationship in the supply chain~\cite{cyclonedx,ntia-sbom}.
A primary tool for transparency is the Software Bill of Materials (SBOM)~\cite{cisa-sbom}.
A SBOM is an inventory of all the components that that make up a software product.
SBOM promotes transparency by tracking component metadata, enabling mapping to other sources of information, and tying the metadata to software as it moves down the supply chain and is deployed~\cite{ntia-minimum-sbom}. As security vulnerabilities are discovered in these components, applications that depend on them can be quickly and reliably tracked and updated to newer, hardened versions.

An SBOM provides a foundation for additional capabilities that enhance software supply chain security, \eg Component Analysis. Component Analysis is a function within an overall Cyber Supply Chain Risk Management (C-SCRM) framework as it helps to understand and manage risks that these components may present to the missions they support. As the software supply chain extends due to the increasing system complexities, it is increasingly important to understand, evaluate, and manage the risk that various components in the supply chain may present in addition to its function. For example, the criticality of artifacts and processes could be used to determine data dependency between components. Building on a set of multidisciplinary publications, standards, and guidelines~\cite{ISO/IEC_27001_nodate,boyens_supply_2015,ISO/IEC_20243_nodate-1,ross_systems_2018,ISO/IEC_27002_nodate-2,joint_task_2013,ISO/IEC_27036_nodate}, Paulsen \etal~\cite{paulsen_criticality_2018} proposed a Criticality Analysis Process Model that prioritizes programs, systems, and components based on their importance to the goals of an organization and the impact that their inadequate operation or loss may present to those goals.

Some techniques provide SBOM-similar information to users. 
For example, Sigstore's~\cite{newman2022sigstore} transparency log gives users the ability to view information about artifacts and operations used to create them;
npm-audit~\cite{npm_docs_npm-audit_nodate} gives users a way to visualise dependencies; and
git commit signing~\cite{git-commit-signing} provides authorship information.
Several other tools automate information collection and assurance across the supply chain.

The collection of this information founds trust within the supply chain.
By understanding how artifacts, operations, and actors interact, members of the supply chain can trace dependencies back to a trusted root source.
This enables the flow of trust between entities in the supply chain.
The promotion of validity ensures that the interactions between supply chain elements are certifiable --- increasing trust.

\subsection{Promoting Validity}

Establishing trust entails providing security at every step, which incrementally evolves, each built on the preceding to provide incremental confidence. While end-users know what components make up their software, can they trust what the component says it is? Furthermore, believe the integrity of the executable received is intact? 
High reliance on open and closed source packages has decreased the confidence that systems only do what they are intended for. Manual code review helps to identify vulnerabilities, but since some software are distributed as prebuilt binaries, it is less effective to manually review the individual source code for malicious flaws.

Lamb \etal~\cite{lamb_reproducible_2022} and Goswami \etal~\cite{goswami_investigating_2020} looked at the problem of establishing trust in build artifacts by comparing build outputs from multiple independent builders~\cite{knightExperimentalEvaluationAssumption1986, mckeemanDifferentialTestingSoftware1998}. 
This way, the user can verify that the received binaries are identical to other builds. The reproducible build approach is a countermeasure solution to attacks that could compromise the executables at build time, where changes are essentially invisible to its original authors and users alike. 
Code signing and verification is an integral part of ensuring that software is from an original publisher. It ensures that the final published software is intact and contains no tampering from unauthorized parties. Sigstore~\cite{newman2022sigstore} improves the integrity of the software supply chain by combining various technologies to provide an automated approach for developers to digitally sign artifacts and for users to verify the artifacts in Sigstore’s transparency log - a public, tamper-proof ledger of signatures. This mapping of artifacts to verifiable identities establishes trust that they are tamper-free. Gitsign implements keyless Sigstore to sign git commits with a valid OpenID Connect identity which overcomes the challenges associated with using GPG keys in signing git commits and consequently improving the overall trust in open-source projects.
Many point solutions have been implemented~\cite{reproducible-builds,tuf,git-commit-signing,vu_lastpymile_2021} to ensure that individual supply chain actions are not altered.
Torres-Arias \etal~\cite{torres-arias_-toto_2019} implemented a holistic approach that enforces the integrity of a software supply chain by gathering cryptographically verifiable information about the chain itself and verify that each step action of a supply chain is not tampered with. 
This approach ensures end-to-end verification and confirms that tampering does not occur in between steps in the software supply chain between the development and the publication of the software.

Account takeover attacks place the account owner and anything the account has access to at risk. Multi-factor Authentication is highly recommended for actors in the supply chain. Although all downstream systems that depend on the affected code are impacted, no solution provides the security posture of actors (e.g. maintainer, developer) in a supply chain to enable software consumers to make risk-based security decisions. Will it have more cons than pros? 
Several C-SCRM practices for systems and organizations have been recommended to mitigate attacks due to credential compromise. Building on existing FIPS 200 standards~\cite{technology_minimum_2006}, Boyens \etal~\cite{boyens_cybersecurity_2022} added that information system access should be limited to only the necessary type and duration and monitored for cybersecurity supply chain impact. They also expanded the Awareness and Training control of FIPS 200 to include providing C-SCRM awareness and training to individuals at all levels within the enterprise as well as suppliers, developers, system integrators, external system service providers, and other information technology (IT)- or operational technology (OT)-related service providers to ensure that the personnel who interact with an enterprise’s supply chains receive the training as appropriate.

\subsection{Promoting Separation}

Several techniques currently exist for promoting separation between components in a supply chain.
We distinguish between those techniques that apply to internal components and those that apply to external links.
Internal-focused techniques ensure compartmentalization between artifacts, operations, and actors of a single link.
External-focused techniques mitigate the risk associated with relying on other links in the supply chain.

First, container and virtual machine (VM) based methods separate internal operations, artifacts, and actors~\cite{ensor_shifting_2021, microsoft_3_2021, solarwinds_setting_2021}. 
Successful build systems are highly automated~\cite{cloud_native_computing_foundation_software_2021}. 
In these systems, automated workers can behave as actors by managing operations and artifacts.
In these cases, best practice is the creation of ephemeral and task-specific workers~\cite{cloud_native_computing_foundation_software_2021}.
This prevents any unnecessary crossover between logically different operations and actors - reducing the risk associated with internal connections.
Additionally, the creation of compartmentalized containers and VM instances reduces the risk associated with connected artifacts.
As an example, BreakApp~\cite{vasilakis_breakapp_2018} assists in spawning compartmentalized modules to complete tasks during development.
By systematically creating separated build systems, attackers cannot effectively propagate vulnerabilities throughout the supply chain.

Second, actors adopt systems which provide separation from external supply chain links. 
For example, version locking and mirroring techniques add a layer of security to external sources~\cite{cloud_native_computing_foundation_software_2021, microsoft_3_2021, ellison2010evaluating}.
Version locking ensures that a link includes a particular version of an upstream component.
The use of a constant dependency version ensures that malicious changes upstream do not automatically propagate to downstream links.
A weakness to version locking is the reliance on actors to accurately set and manage version numbers.
The failure to update versions may prevent updates that remove vulnerabilities.
In the same manner, prematurely updating to a compromised version defeats the purpose of version locking.
Mirroring acts in a similar manner to version locking. 
Organizations create private package feeds to mitigate the risk of pulling dependencies from public sources~\cite{microsoft_3_2021,winters2020software}.
This gives organizations more control over the import of packages into their software projects.
Once again, this relies on including the correct packages in private feeds.

\section{Mapping Embodiments to Security Properties} \label{sec:embodiments}

In this section, we use case studies to demonstrate how each security property from~\cref{sec:properties} can be embodied in practice.
We specifically consider efforts to secure supply chains through
  (1) package repositories,
  (2) development environments,
  (3) end-to-end solutions, and
  (4) security frameworks.

\subsection{Package Repositories}

Package and dependency managers such as npm for JavaScript, PyPI for Python, or RubyGems for Ruby have encouraged code reuse between packages. Consequentially package managers have become a vital part of software supply chains.
Attackers have begun exploiting the weak links in these ecosystems to distribute malware.
Several techniques have been put in place to detect and mitigate package manager ecosystem attacks.
We examine Npm as a case study to discuss the techniques that have been adopted to improve the security properties defined in~\cref{sec:properties}.

Npm is the package manager for the popular Node.js JavaScript platform.
Since Node.js is widely used for back-end software, Npm is widely used in security-sensitive contexts.

\textbf{Transparency:}
To use Npm as a package manager, a software engineer must declare the packages on which they depend in a file called \code{package.json}.
The command \code{npm install} command can then installs these packages.
The \code{package.json} file is a ``primitive SBOM'' since it contains almost all the required fields of one (as recommended 
by the NTIA~\cite{ntia-minimum-sbom}). SBOM's in general, provide an example of the Transparency property in practice.

As one application of this primitive SBOM, Npm offers a tool called \emph{npm-audit}~\cite{npm_docs_npm-audit_nodate} to audit package security and stability. Npm-audit assesses the dependency description of a project against the default registry for possible security vulnerabilities and calculates the impact and appropriate remediation if any are found. The audit process entails scanning the package.json and package-lock.json file to build the dependency tree, then comparing the packages from the dependency tree to a database of known vulnerabilities. If any vulnerabilities are found, an alert with the impact and appropriate remediation will be shown. Although security audits help you protect package users by enabling you to find and fix known vulnerabilities in dependencies that could cause data loss, service outages, and unauthorized access to sensitive information, recent researchers have started doubting the reliability of this tool~\cite{liu_demystifying_2022, wyss_what_2022}.

\textbf{Validity:}
Npm has recorded several incidents on the registry where npm accounts are hijacked by malicious actors and the use of the access to infiltrate packages the compromised accounts have access to. Following the unprecedented series of account takeovers resulting from the compromise of developer accounts without 2FA enabled ~\cite{cisa_malware_nodate, noauthor_security_2021}, npm has introduced even more security enhancements. 
An extra layer of security of verifying all npm account login, and enrolling maintainers of top packages into a mandatory 2FA have been added to help prevent common accounts takeover attacks, such as credential stuffing~\cite{thomas_protecting_nodate}, which utilize a user’s compromised and reused password.

\textbf{Separation:}
 Npm provides scopes to safeguard a diverse set of package names by restricting the package’s namespace to an organization or user~\cite{npm_scope}. This means that one does not have to worry about someone taking a package name ahead of time and only the user is allowed to publish packages under that scope on the public registry, which hardens the process against compromise as an attacker would have to compromise that npmjs.org registry account to take over the package. Scopes can be associated with a given registry which ensures that all requests for packages under the scope will be routed to the given registry. For instance, at login the \textit{myorg} scope can be linked to \textit{http://registry.myorg.com} with the command npm login:

\begin{lstlisting}[language=bash]
  $ npm login --scope=@myorg 
       --registry=http://registry.myorg.com
\end{lstlisting}

This command will ensure that any request bound to the \textit{myorg} scope is sent to the \textit{http://registry.myorg.com} registry. Scopes mitigate the dependency confusion risks where an Internal package name is claimed by an attacker on the public registry.

The use of npm proxy~\cite{npm_proxy} is a common practice for improved npm security. The internal registry can be configured to take precedence over the public registry to help protect against installing the wrong or malicious package from public registries. For example, an attacker might publish a malicious package to the public repository with the same name as a package hosted on a private registry but with a higher semantic version. In the case where a custom setting for an internal registry is omitted, the package manager would default to the public registry and download the latest (malicious) packages from there. Configuring the proxy to never allow an upstream request to the public registries protects against fetching arbitrary packages in place of the legitimate package.

\subsection{Development Environments}

Collaborative development environments like GitHub and Bitbucket host millions of open- and closed-source projects.
The meritocratic premise of open-source software~\cite{raymond1999cathedral}, which allows contributions from anyone, is an opportunity for the introduction of vulnerabilities through ``hypocrite commits''~\cite{wu2021feasibility}.
Hence, such platforms have taken steps to promote the security of the projects they host.
We align these steps with our proposed security properties in a case study of GitHub.

    
\textbf{Transparency:}
In a package context such as Npm, the package manager can mandate the use of a \code{package.json} file. This file can be leveraged for further analysis.
In contrast, in the source code context, the supply chain for a given project is harder to discover.
GitHub~\cite{noauthor_secure_2020} identified unpatched software as the major threat to supply chain security and has provided capabilities of Software Composition Analysis (SCA) to determine dependencies in use, discover vulnerabilities in the dependencies, and effect patches. These capabilities are provided by dependency graph, dependabot alerts, and dependabot security updates feature and are recommended to organizations to help secure their repositories against supply chain threats~\cite{github_secure_your_nodate}.

The dependency graph is a summary of manifests and lock files that shows the dependencies and dependents of your repository. When a pull request containing changes to dependencies that target the default branch is created, GitHub uses the dependency graph to add dependency reviews to the pull request. These indicate whether the dependencies contain vulnerabilities and, if so, the version of the dependency in which the vulnerability was fixed. Dependabot alerts rely on the dependency graph and GitHub advisory database to alert developers when a repository is affected by a newly discovered vulnerability. This enables organizations and open-source projects to stay up to date on security vulnerabilities, and information. Dependabot security updates make it easier to mitigate this vulnerability within repositories by automatically raising pull requests to update a software dependency to the minimum version that resolves a known vulnerability. These dependabot features provide automation to the hard work of dependency management and patching. However, the extent to which the dependency update bot reduces update suspicion and notification fatigue remain questionable ~\cite{he_automating_2022, noauthor_enable_2022}.

To eliminate the security risk posed by the late detection of vulnerabilities, GitHub~\cite{about_code_scanning} developed a feature --- static analysis, powered by CodeQL, that runs queries against codebases to identify potential security vulnerabilities.

\textbf{Validity:}
To secure build systems against build process attacks, Github Actions (CI/CD tool for GitHub) is designed to ensure precise and repeatable build steps and that each build starts in a new environment to reduce the likelihood of attackers persisting in a build environment~\cite{github_secure_build}.
GitHub builds on top of the git version control system, which has a feature to enable developers to validate that commits are coming from an identified, trusted source while using other people's work.
Specifically, git supports signing and verifying commits and tags using GPG~\cite{git-commit-signing}.

\textbf{Separation:}
Platforms such as GitHub allow software developers to create arbitrarily many independent repositories (within reason).
Links between these repositories (\eg by adding a dependency or a git sub-module) are at the developer's discretion.
This decision improves the separation between links in the supply chain.
However, we note that the common practice of \emph{vendoring} --- copy-pasting a dependency verbatim into another codebase --- degrades separation.

\subsection{End-to-end Solutions}

Some researchers have focused on developing fully-fledged systems to mitigate supply chain attacks.
Solutions such as in-toto~\cite{torres-arias_-toto_2019} and Sigstore~\cite{newman2022sigstore} are excellent examples and have been integrated across vendors to secure software supply chains~\cite{noauthor_-toto_nodate}.
We examine in-toto as a case study for the implementation of security properties proposed in \cref{sec:properties}.

In-toto ensures the security of the software supply chain by gathering cryptographically verifiable evidence --- called \emph{link metadata} --- about entities in the chain.
\emph{Link metadata} is a signed statement each actor in the supply chain emits to describe relevant operations, artifacts, and connections.
For example, this statement may include information such as files used, files produced, building processes, or even environment variables.
\emph{Link metadata} is collected throughout the supply chain and is delivered alongside the final product.
Verifiers can compare \emph{link metadata} with a \emph{layout} (created by an actor who dictates policy for the supply chain) describing the intended steps in the supply chain.

\textbf{Transparency:}
By disaggregating the supply chain into small, individual claims from each actor, in-toto provides strong transparency guarantees for artifacts (they are tracked as materials and products) and operations (they are described in a layout file). 
However, it does not provide transparency for actors in the supply chain.
Although actors will create signed statements, in-toto does not have a method to tie their keys to identities (even pseudonyms). 
While this is not exclusive to the ``holistic'' approach of the system, it may be possible to extend the design to include known identifiers using approaches such as verifiable credentials~\cite{vc} or distributed identity providers~\cite{did}.

\textbf{Validity:}
Regarding validity, in-toto provides guarantees for artifact integrity (by means of hashing each artifact) and operation integrity (by means of a layout policy), yet it does not provide actor authentication.
Even though the signing keys prevent a malicious takeover of operations carried out by an actor, it does not protect against other attack vectors (\eg account take over). 
It is possible that author validity guarantees may be achieved by combining existing mechanisms (\eg 2FA) into the signing flow for link metadata. 
Likewise, actors provide additional proof they are a well-known, reputable actor in the chain through techniques like DiD and VC described above.

\textbf{Separation:}
Lastly, in-toto provides strong Separation guarantees for operations and actors, but not for artifacts.
This is because artifact separation is typically achieved by underlying mechanisms (\eg a container runtime may sandbox a build process). 
Given this, it may be possible to provide stronger security guarantees by adopting hardened runtimes within solutions like in-toto. 
One example of this is SLSA's extension to in-toto links~\cite{slsa}, that can help communicate information if the build is hermetic, or if all the artifacts used in the build were required by the build.

\subsection{Security Frameworks}

Researchers and industry have proposed multiple end-to-end security frameworks to simplify the process of securing a supply chain. These frameworks recommend practices, tooling options and design considerations to ensure the integrity of artifacts in the software supply chain. We consider three frameworks --- Microsoft's Supply Chain Integrity Model (SCIM), Google’s Supply-chain Levels for Software Artifacts (SLSA) and The Cloud Native Computing Foundation's (CNCF) Software Supply Chain Best Practices. 

First, SCIM~\cite{scim} specifies how the artifact verification process should function across a supply chain.
The framework provides a standard for the artifact data model and exchange format (SCIM-Evidence), the policy used in evaluating these artifacts (SCIM-Policy), and the service that stores both the evidence and the policy (SCIM-Store).
This enables the smooth flow of evidence (\eg bills of materials, build information, etc.) between links in the supply chain.

Second, SLSA~\cite{slsa} provides four levels of assurance (\ie SLSA 1-4) to describe the security posture of a supply chain, similar to the capability maturity model (CMM) for software development~\cite{humphrey_characterizing_1988}.
The framework starts with basic security at SLSA 1 and adds requirements at each level until the final level, SLSA 4.
Each level specifically documents source, build, provenance, and common requirements. 

Finally, the CNCF~\cite{cloud_native_computing_foundation_software_2021} framework provides a five-stage methodology for securing supply chains.
These stages include securing the source code, materials, build pipelines, artifacts, and deployments of a supply chain.

In \cref{tab:frameworks}, we compare how the three frameworks cover our security properties. 
SCIM provides substantially less coverage than SLSA 4 and CNCF.
Although SLSA 4 and CNCF cover our security properties better, they may be difficult to implement\footnote{SLSA 4 lists 20 requirements; the CNCF proposes nearly 60 requirements.}.

\begin{table}[ht!]

\caption{
  Comparison of how SCIM~\cite{scim}, SLSA~\cite{slsa}, and CNCF~\cite{cloud_native_computing_foundation_software_2021} promote security properties.
  }
\begin{tabular}{|cc|c|c|c|}
    \hline
    \multicolumn{2}{|c|}{\textbf{Frameworks}} & SCIM & SLSA 4 & CNCF \\
      
    \hline
    \hline

    \multirow{3}{*}{\textbf{Transparency}}
    & \footnotesize Artifacts & \checkmark & \checkmark & \checkmark\\
    \cline{2-5}
    & \footnotesize Operations & \checkmark & \checkmark & \checkmark\\
    \cline{2-5}
    & \footnotesize Actors &  & \checkmark & \checkmark\\

    \hline
    \hline

    \multirow{3}{*}{\textbf{Validity}}
    & \footnotesize Artifacts & \checkmark & \checkmark & \checkmark\\
    \cline{2-5}
    & \footnotesize Operations & \checkmark & \checkmark & \checkmark\\
    \cline{2-5}
    & \footnotesize Actors &  & \checkmark & \checkmark\\

    \hline
    \hline
    
    \multirow{3}{*}{\textbf{Separation}}
    & \footnotesize Artifacts &  & \checkmark & \checkmark\\
    \cline{2-5}
    & \footnotesize Operations &  & \checkmark & \checkmark\\
    \cline{2-5}
    & \footnotesize Actors &  &  & \checkmark\\
    
    \hline

\end{tabular}
\label{tab:frameworks}
\end{table}


\textbf{Transparency:}
\textit{SCIM} improves the supply chain's transparency by providing principle for conveying evidence about the artifact. For example, information about the sub-components of an artifact, how the artifact was created, and defects identified in the artifact.
\textit{SLSA} promotes transparency by demanding provenance attestation files at the lowest compliance level. These files contain build metadata which inform users about the artifacts they use. 
The \textit{CNCF} framework proposes several practices to increase transparency in the supply chain, such as SBOMs and dependency analysis. 

\begin{table*}
\caption{
  Proposed and practical techniques, related to security properties (columns) and aspects of the definition of a supply chain (sub-columns).
  Note the emphasis on \emph{artifacts} and the under-emphasis on \emph{operations} and \emph{actors}.
  We believe this gap represents an opportunity for security research.
  We suggest that vetting \emph{operations} may require applied cryptography and improved hardware root-of-trust.
  Meanwhile, addressing security flaws related to \emph{actors} will require accounting for human and organizational factors, a longstanding challenge in cybersecurity.
  }
\begin{tabular}{|c||c|c|c||c|c|c||c|c|c||}
    \hline
    \multirow{2}{*}{\textbf{Techniques}}
     & \multicolumn{3}{c||}{\textbf{Transparency}} 
     & \multicolumn{3}{c||}{\textbf{Validity}}
     & \multicolumn{3}{c||}{\textbf{Separation}} \\
    
    & \footnotesize Artifacts & \footnotesize Operations & \footnotesize Actors
    & \footnotesize Artifacts & \footnotesize Operations & \footnotesize Actors
    & \footnotesize Artifacts & \footnotesize Operations & \footnotesize Actors \\
    
    \hline \hline

    SBOM & \checkmark & \checkmark &  &  &   &   &   &   &\\
    \hline
    npm-audit~\cite{npm_docs_npm-audit_nodate} & \checkmark &   &  & \checkmark &   &   &   &   &\\
    \hline
    Code scanning~\cite{about_code_scanning} & \checkmark &  &   & \checkmark  &   &   &   &   &\\
    \hline
    Dependabot features~\cite{github_secure_your_nodate}  & \checkmark  &   &   & \checkmark &   &   &   &   &\\
    \hline
    GitHub Actions~\cite{github_secure_build} &   & \checkmark &   & \checkmark  & \checkmark &   &   & \checkmark &\\
    \hline
    Git Commit Signing~\cite{git-commit-signing} &  &   & \checkmark & \checkmark &   &   &   &   & \\
    \hline
    Scope~\cite{npm_scope} &   &   &   & \checkmark &    &   & \checkmark  &   & \checkmark\\
    \hline
    Multi-Factor Authentication &   &   &   &   &   & \checkmark  &   &   &\\
    \hline
    In-toto~\cite{torres-arias_-toto_2019} & \checkmark  & \checkmark &   & \checkmark & \checkmark  &   &  & \checkmark  & \checkmark\\
    \hline
    Containerization &  &   &   &   &   &   & \checkmark & \checkmark & \checkmark\\
    \hline
    Version Locking &  &   &   &   &   &   & \checkmark &  &\\
    \hline
    Sigstore~\cite{newman2022sigstore} & \checkmark & \checkmark & \checkmark & \checkmark & \checkmark &   &  &  & \\
    \hline
    Mirroring and Proxies~\cite{npm_proxy} & \checkmark &   &   & \checkmark &   &   & \checkmark & \checkmark &\\
    \hline
    
\end{tabular}
\label{tab:techniques_to_principles}
\end{table*}

\textbf{Validity:}
\textit{SCIM} improves the validity of the supply chain by providing principles for conveying the evidence of an artifact that allows it to be verified. For example, providing the principles for conveying the cryptographic hash of a software allows the software to be verified by the consumer.
\textit{SLSA} 2 to 4 generates validity requirements for implementers. At these levels, SLSA demands version control, provenance integrity, auditability, and a two-party review of all changes to the artifact. Such specifications ensure that artifacts and operations within the supply chain are tamper-free. 
The \textit{CNCF} framework recommends a combination of cryptographic attestation and verification at each stage of the supply chain (verification by reproducibility, Multi-factor authentication, and signature validation at every step, amongst others) to ensure integrity.

\textbf{Separation:}
The \textit{SCIM} framework does not provide specific methods to increase separation in the supply chain.
\textit{SLSA} 3 requires isolated builds in ephemeral environments (dedicated resources for that particular build) and SLSA 4 requires a hermetic build process. These requirements provide strong protection against cross-build contamination attacks~\cite{slsa}. 
Also, the \textit{CNCF} framework recommends several practices that embody separation, including maintaining controlled source code, build, and artifact environments. Controlled environments are created by adopting policies such as branch protection rules, ephemeral build workers, ephemeral certificates, pipeline orchestration, minimal network connectivity, build worker segregation, and artifact access rules.

\section{Discussion} \label{sec:discussion}

In~\cref{sec:approaches} and~\cref{sec:embodiments} we related current proposals and embodiments to the security properties from~\cref{sec:properties}.
We believe this demonstrates the usefulness of the properties we have proposed.
As a last question, we consider whether these approaches are comprehensive with respect to software supply chains.
Recall the definition of a software supply chain given in~\cref{sec:background}: \emph{a collection of systems, devices, and people which produce a final software product} \cref{fig:supply_chain}.

We mapped each technique described in~\cref{sec:approaches} and~\cref{sec:embodiments} to the corresponding security properties and the relevant aspect(s) of the software supply chain.
\cref{tab:techniques_to_principles} presents our results.
We observe that most of the studied approaches are focused on artifacts.
Further research is needed to assess the generalizability of this claim beyond the cases we examined.
If true, we acknowledge that artifacts are the appropriate primary focus for security --- when deployed, software systems consist of interacting software artifacts, and cybersecurity vulnerabilities consist of exploiting these interactions.
However, we suggest that there are many opportunities to improve the handling of operations and actors in secure software supply chains.

\section{Conclusion}

In this paper, we proposed desired properties of a secure software supply chain and systematized current design patterns and practices. 
We analyzed security frameworks and several real-world techniques and mapped these techniques according to the corresponding principles they promote. 
Our analysis showed how current practices embody the proposed security properties and apply to components of the software supply chain.
We intend for this systematization to serve as a reference and guide for those seeking to build frameworks and improve the security of software supply chains.

\section*{Acknowledgments}
We acknowledge support from Cisco as well as NSF award \#2229740. 
We thank the reviewers and our shepherd, Asra Ali, for their thoughtful critiques.


\raggedbottom
\pagebreak

\ifBIBLATEX
  \setlength\bibitemsep{0.00\itemsep} 
  \printbibliography
\else
  \bibliographystyle{abbrv}
  \bibliography{refs/sok.bib, refs/jamie-bib.bib, refs/pose-santiago-bibdata.bib, refs/PurdueDualityLab.bib, refs/Chinenye-SoK-bib.bib}
\fi

\end{document}